\documentclass[letterpaper]{article}
\usepackage{aaai}
\usepackage{times}
\usepackage{helvet,amsthm}
\usepackage{courier,url}
\usepackage{graphicx,pifont,multirow,diagbox}
\usepackage{color,booktabs}
\usepackage{float,makecell}
\usepackage{verbatim}
\usepackage[flushleft]{threeparttable}

\usepackage[most]{tcolorbox}
\usepackage{lipsum}

\usepackage{subfigure}

\usepackage{enumitem}
\usepackage{array}
\newcolumntype{H}{>{\setbox0=\hbox\bgroup}c<{\egroup}@{}}

\newcommand{\m}{GraphUIL}

\theoremstyle{problem}
\newtheorem{problem}{Problem}
 
\begin{document}
%
\title{Deep Network Embedding for User Identity Linkage}
\title{Multi-stage Aggregation Neural Networks for User Identity Linkage}
\title{Deep Local and Global Network Embedding for User Identity Linkage}
\title{Graph Neural Networks for User Identity Linkage}

\author{Wen Zhang\thanks{Equal contribution}, Kai Shu\footnotemark[1], Huan Liu, and Yalin Wang \vspace{0.1in} \\
Computer Science and Engineering, Arizona State University, Tempe, 85281, USA\\
\{wzhan139, kai.shu, huan.liu, ylwang\}@asu.edu
}
\maketitle

\begin{abstract}

The increasing popularity and diversity of social media sites
has encouraged more and more people to participate in multiple online social networks to enjoy their services. Each user may create a user identity to represent his or her unique public figure in every social network. User identity linkage across online social networks is an emerging task and has attracted increasing attention, which could potentially impact various domains such as recommendations and link predictions. The majority of existing work focuses on mining network proximity or user profile data for discovering user identity linkages.

With the recent advancements in graph neural networks (GNNs), it provides great potential to advance user identity linkage since users are connected in social graphs, and learning latent factors of users and items is the key. However, predicting user identity linkages based on GNNs faces challenges. For example, the user social graphs encode both \textit{local} structure such as users' neighborhood signals, and \textit{global} structure with community properties. To address these challenges simultaneously, in this paper, we present a novel graph neural network framework ({\m}) for user identity linkage. In particular, we provide a principled approach to jointly capture local and global information in the user-user social graph and propose the framework {\m}, which jointly learning user representations for user identity linkage. Extensive experiments on real-world datasets demonstrate the effectiveness of the proposed framework.

\end{abstract}

\section{Introduction}


With the prosperity of online social networks, people's social activities are closely related to others with different social purposes such as information seeking/sharing and social connection maintenance. Due to the diverse functionalities provided by different social media sites, one person might join multiple sites to serve their different needs. For example, one user may use Twitter~\footnote{https://twitter.com/?lang=en} to publish personal opinions on political events while, on the other side, constantly shares travel photos and his/her leisure life story only on Instagram~\footnote{https://www.instagram.com/?hl=en}.
One of the fundamental tasks in social networks mining is to effectively identify the same anonymous individuals across platforms, which is also termed as \textit{user identity linkage}~\cite{shu2017user}. By linking user identities across social networks, we can obtain a comprehensive view of a person's social characters and interests, which is crucial to cross-network information diffusion~\cite{peng2013predicting} and cross-domain recommendation~\cite{huang2016social}. Thus, it is important to identify user identities across social networks.

Existing approaches to link user identities mainly focus on extracting discriminative features such as hand-crafted user profile features~\cite{zafarani2015user} , or building effective latent models such as network embeddings~\cite{zhou2018structure,man2016predict}. However, these methods may face the following challenges: (1) online social media data user generated is generally noisy and biased across different platforms~\cite{tang2014mining}. For example, the social networks of users in Twitter may be related to serious political like-minded people, while in Facebook they may include more personal known friends; (2) user attributes are often incomplete and missing in social networks~\cite{narayanan2010myths}. For example, usually only a part of individual social characters may be expressed and many users tend to hide their identities in different online social networks to preserve their privacy. One way to build an effective user identity linkage system is to learn user representations only using user social networks with handling the noisy and bias issues.

\begin{figure*}[!]
\center
    \includegraphics[width=0.85\linewidth]{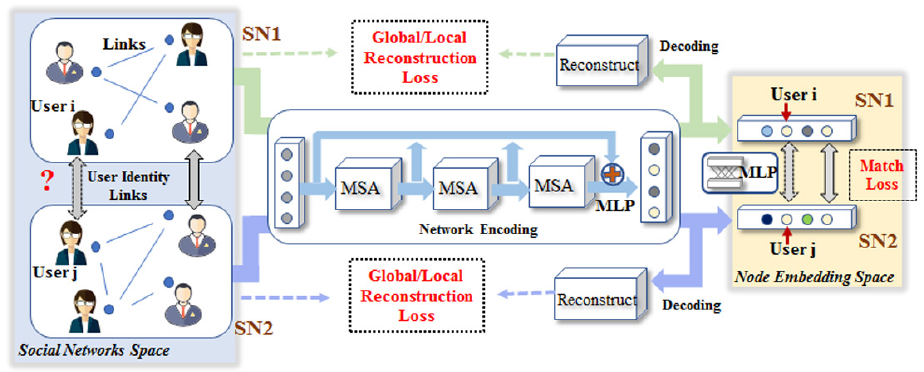}
    \caption{The pipeline of GraphUIL. It is a semi-supervised learning model. Given two social networks (SN1 and SN2), a network encoding is performed respectively for each of them with stacked MSA layers and long skip connections. The encoded node features are applied for network reconstruction (Decoding) and meanwhile feed to an MLP network for matching user identity across networks. }
    \label{fig:lossrecord}
\vspace{-1em}
\end{figure*}

Recently, Graph Neural Networks (GNNs) have shown promising results on learning node representations on network structures~\cite{kipf2016semi,defferrard2016convolutional}. Their main idea is to aggregate feature information from \textit{local} graph neighborhoods
using neural networks. Meanwhile, node information can be
propagated through a graph after transformation and aggregation to obtain a \textit{global} view. For example, Tyler \textit{et al.} utilize graph neural networks for signed link prediction~\cite{derr2018signed}, and Ma \textit{et al.} propose to learn network representations through GNNs in streaming networks~\cite{ma2018dynamic}.  Despite the success of existing GNN based methods, most of them focus on handling single network data and may not be applicable for user identity linkage task. In addition to model the network structures across sites, the learned node representations in latent space should also preserve the linkage information explicitly.

Therefore, in this paper, we study the novel problem of learning user representations through graph neural networks for user identity linkage. In essence, we investigate the following challenges: 1) how to effectively learn user representations across social networks only using social graph information, and 2) how to mathematically formulate user identity linkage problem through graph neural networks so as to improve prediction performance.  To tackle these challenges, we propose a graph neural network framework for user identity linkage ({\m}), which can i) learn latent user representations to preserve both global and local network structures simultaneously; and ii) build a non-linear cross-platform mapping kernel to predict user identity linkages. The main contributions are summarized as follows.

\begin{itemize}
    \item We provide a principled way to capture local and global structure information for learning node representations;
    \item We propose a novel framework {\m} based on graph neural networks, which can encode local and global network signals for user identity linkages across social networks; and
    \item We conduct experiments on real-world datasets to demonstrate the effectiveness of the proposed framework for linking user identities.
\end{itemize}

\section{Problem Formulation} \label{sec:problem}
A social network graph is defined as $\mathcal{G}=\{\mathcal{V},\mathcal{E}\}$, where $\mathcal{V}=\{v_i|i=1,...,N\}$ represents a set of $n$ nodes (users) in the graph $\mathcal{G}$ and $\mathcal{E}=\{e_{i,j}|(i,j)=1,...,N\}$ represents the edges (association between users). On each node $v_i$, we define a node feature $x_i$ which can be the user profile, content or graph structure properties on that node. Each edge $e_{i,j}$ is associated with a weight $a_{i,j}\in \mathbf{R}$. Here, we let $a_{i,j}=1$ when node $v_i$ and $v_j$ are linked, otherwise $a_{i,j}=0$. The matrix $A=\{a_{i,j}\}\in \mathbf{R}^{N\times N}$ is called the adjacent matrix of graph $\mathcal{G}$.

Suppose we have multiple social networks (two networks in this study) and each network contains a large number of nodes, we intend to find those nodes which are the same user identity across networks, namely user identity linkage predictions. Here, we only rely on the network topology to learn such a prediction. The most challenge part of this task is the irregular graph patterns across networks. For example, given two social networks, $\mathcal{G}^{(1)}$ is designed for people to share photos, e.g. Instagram, and the other $\mathcal{G}^{(2)}$ is designed to share short news and personal messages, e.g. Twitter, the same user might exhibit different social states and actions in these two social networks. Besides, most social networks have sparse graph structures and a large number of users has similar local graph patterns, i.e. followed by the same group of friends. Our problem is formally defined as below.
\begin{problem}
Given two social networks $\mathcal{G}^{(1)}=(\mathcal{V}^{(1)},\mathcal{E}^{(1)})$ and $\mathcal{G}^{(2)}=(\mathcal{V}^{(2)},\mathcal{E}^{(2)})$, the task is to predict whether two user identities $v^1\in\mathcal{V}^{(1)}$ and $v^2\in\mathcal{V}^{(2)}$ belong to the same real person, i.e., $F:\mathcal{V}^{(1)}\times \mathcal{V}^{(2)}\Longrightarrow\{0,1\}$.
\end{problem}

\section{{\m}: Graph Neural Networks for User Identity Linkage}\label{sec:framework}
In this section, we introduce the details of the proposed {\m} framework.
We first encode graph topology of social networks into node features. This process of feature learning is a task of network representation learning, called node embeddings. The purpose of such embedding is to project the network structure to the low dimensional node space, in which process both global and local graph connection patterns are preserved such that the reconstructed networks based on the learned node features are close to the original networks. In this work, we design a deep graph model to learn node embeddings for the large social networks and build a non-linear mapping of nodes which have the same user identity across networks. Such semi-supervise learning strategy helps us to predict the user identity linkage effectively in the real cases.


\begin{figure}[t]
\center
    \includegraphics[width=0.65\linewidth]{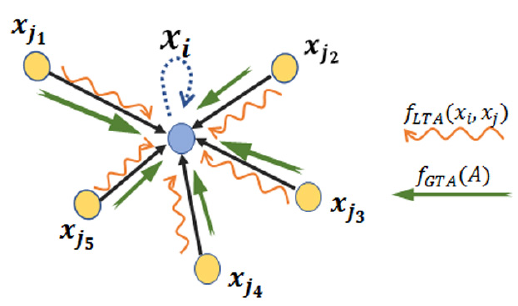}
    \caption{The multi-stage aggregation layer (MSA).}
    \label{fig:MSA}
\vspace{-1em}
\end{figure}

\subsection{Network Embedding Through Node Propagation}

The graph representation learning should capture the global graph topology as well as the local neighborhood graph patterns to get a thorough encoding of network structures. The encoded network structures are generally translated as the node information. In other words, the network embedding transfers edge links to node features. Here, we propose to use graph propagation mechanisms to achieve this goal. The node information is updated based on the aggregated information of its neighborhood nodes. We use the 1-hop aggregator, which relies on the directly linked nodes, as the unit element in our framework and by stacking multiple such aggregators, the larger neighborhood areas are taken into account. 
Next, we mathematically explain the network aggregation layer with deep network structures. An aggregator function $Agg(.)$ on a node $v$ defines a specific pattern of aggregating the node information of its 1-hop neighborhood. The updated node features after aggregation can be expressed as: 
\begin{equation}\label{eq:general_agg}
   x^{'}_v=\delta(W\cdot Agg(x_v,\{x_u\})),
\end{equation}
where $\{x_u,\forall u\in \mathcal{N}(v)\}$ is the 1-hop neighbourhood node features. $W$ is the linear transformation weight matrix. We associate the aggregation function with the graph adjacent matrix $A$, and generalize Eq.~\ref{eq:general_agg} to the node feature matrix $X$,
\begin{equation}
    X^{'}=\delta(AGG(A,X)\cdot W).
\end{equation}
It can be realized that an aggregation pattern defines how we accumulate neighborhood information by learning a set of aggregation weights on each linked edges and a pooling function such as mean, sum or max-pooling. Thus, having different aggregation strategy with respect to graph topology $A$ and original node information $X$, we obtain various node embedding results by updating $X^{'}$ iteratively. In this study, we designed a multi-stage aggregator (Fig.~\ref{fig:MSA}) which considers the global and local network structures simultaneously.

\subsubsection{Global-Topology-Aware Aggregator (GTA-Agg).}
In a social network, a user plays a unique role when interacting with other users. For example, he might be a leader of a virtual chatting group and becomes a hub node with a high node degree and clustering efficiency. To aware some topological properties on a given node, we need to observe the whole network ahead and then assess its node property compared to all its direct or indirect connections. Therefore, the first stage of aggregator should be aware of the global network structures. To this end, we use the spectral-based deep graph convolution which depends on the intrinsic network structures. The convolution works with a spectral representation of the graphs, e.g. graph Laplacian, and learns the spatially localized filters by approximating convolutions defined on the graph Fourier domain. Mathematically, A normalized graph laplacian $L$ is defined as $L=I_N-D^{-\frac{1}{2}}AD^{-\frac{1}{2}}=U\Lambda U^T$, where $D$ is the degree matrix ($D_{ii}=\sum_j A_{ij}$). $\Lambda$ is the diagonal matrix of its eigenvalues and $U$ is the matrix of eigenvector basis. Given a node feature $x\in \mathbf{R}^C$, $U^Tx$ is the graph Fourier transform of $x$. The convolutional operation on this node signal is defined as:
\begin{equation}\label{eq:GraphSpectralConv}
    g_{\theta} \ast x=Ug_\theta U^Tx.
\end{equation}
Fileter $g_\theta=diag(\theta)$ parameterized by $\theta \in \mathbf{R}^N$ is a function of the eigenvalues of $L$, i.e. $g_\theta(\Lambda)$. However, convolution in Eq.~\ref{eq:GraphSpectralConv} is computationally expensive due to the multiplication with high dimensional matrix $U$ and it is a non-spatially localized filters. To solve this problem, it is suggested to use Chebyshev polynomials $T_k(x)$ up to $K^{th}$ order as a truncated expansion to approximate $g_\theta$. The Eq.~\ref{eq:GraphSpectralConv} thus can be reformulated as:
\begin{equation}\label{eq:chebyshev}
    g_{\theta} \ast x\approx \sum^K_{k=0}\theta_kT_k(\tilde{L})x.
\end{equation}
$\tilde{L}=\frac{2}{\lambda_{max}}L-I_N$ and $\lambda_{max}$ is the largest eigenvalue of $L$. Now, $\theta_k$ becomes the Chebyshev coefficients. If we limit $K=1$ and approximate $\lambda_{max}\approx 2$, with the normalized tricks and weak constraints used in ~\cite{kipf2017semi}, Eq.~\ref{eq:chebyshev} simplifies to:
\begin{equation}\label{eq:simple_cheby}
    g_{\theta} \ast x\approx \theta(\tilde{D}^{-\frac{1}{2}}\tilde{A}\tilde{D}^{-\frac{1}{2}})x,
\end{equation}
where $\tilde{A}=A+I_N$ and $\tilde{D}$ is the degree matrix of $\tilde{A}$. We turn Eq.~\ref{eq:simple_cheby} to the matrix multiplication form, for the whole network,
\begin{equation}\label{eq:globalpropagation}
\begin{split}
    X^{'}_{GTA}&=\delta(AGG_{GTA}(A,X)\cdot W_{GTA})\\
    &=\delta(\tilde{D}^{-\frac{1}{2}}\tilde{A}\tilde{D}^{-\frac{1}{2}}\cdot X\cdot W_{GTA}).
\end{split}
\end{equation}
The above equation describes a spectral approach of graph convolution which analogous to a 1-hop node information aggregation. In Eq.~\ref{eq:globalpropagation}, $X\in \mathbf{R}^{N\times C}$ is a graph signal (i.e. node features) with $C$ channels and $X^{'}\in \mathbf{R}^{N\times C^{'}}$ is its convolved signal. The filer parameters matrix $W_{GTA}\in \mathbf{R}^{C\times C^{'}}$ is learned through the backpropagation of deep models.

\subsubsection{Local-Topology-Aware Aggregator (LTA-Agg).}
It is worth noting that the aggregation function defined on Eq.~\ref{eq:globalpropagation} involves degree matrix $D$ which considers the whole graph structure. However, as the large size of users in social networks, in some cases, a sampled dataset for learning contains only a small proportion of users. In other words, the network structure is incomplete. This sampling issue might lead to poor generalization when aggregation is purely based on the adjacent matrix of the sampled data. Therefore, we add another aggregation mechanism aside from the global aggregator and set it as the local attention aggregator. It depends on the signals on the neighborhood to learn aggregation functions and is efficient to compute on the graph spatial domain. This aggregator is more appreciate when a part of social networks observed. The attention mechanism on an aggregator function is formulated as:
\begin{equation}\label{eq:att}
\begin{split}
        X^{'}_{LTA}=&\delta(AGG_{LTA}(A,X)\cdot W_{LTA})\\
        =&\delta(f_{LTA}(A,X)\cdot X\cdot W_{LTA}).
\end{split}
\end{equation}
Here, we use a feedforward neural network to learn attention-based aggregator function $f_{LTA}$, which can be expressed as a matrix $f_{LTA}=A^{'}_{LTA}=\{a^{'}_{i,j}\}\in \mathbf{R}^{N\times N}$, $a^{'}_{i,j}$ is the learnable aggregator weight at a given edge $e_{i,j}$. Suppose we are given two node features, $x_i$ and $x_j$ (all $\in \mathbf{R}^C$), a self-attention from node $v_j$ to $v_i$ is performed on this pair of nodes as:
\begin{equation}
    \phi_{i,j}=\delta_{LTA}(g^T\cdot [W^{'T}x_i\oplus W^{'T}x_j]).
\end{equation}
$W^{'}\in \mathbf{R}^{C\times F}$ is a shared weight matrix for a linear transformation that project all nodes' features to a higher dimensional space to increase the expressive power. $[.\oplus.]$ is the concatenation operator. After projections, a single layer feedforward neural network with activation function $\delta_{att}$, e.g. Relu non-linearity, maps the concatenated node features to a scalar $\phi_{i,j}$. This value indicates how importance of node $v_j$ to $v_i$. The feedforward neural networks is parameterized by the weight vector $g\in \mathbf{R}^{2F}$. We further normalize the neighbourhood attention weights by a softmax function:
\begin{equation}
    a^{'}_{i,j}=\frac{exp(\phi_{i,j})}{\sum_{k\in \mathcal{N}_i}exp(\phi_{i,k})}.
\end{equation}

In the end, we combine aggregation mechanisms in Eq.~\ref{eq:globalpropagation} and Eq.~\ref{eq:att} to obtain our multi-stage aggregation block.
\begin{equation}
\begin{split}
    X^{'}=&\delta(AGG_{GTA}(A,X)\cdot W_{GTA}\\
    &+\lambda\cdot AGG_{LTA}(A,X)\cdot W_{LTA})
\end{split}
\end{equation}
Such aggregation process dynamically learns the 1-hop neighbourhood network patterns. The weights $\lambda$ of linear combination of GTA-Agg and LTA-Agg can be further incorporate in layer parameters $W_{GTA}$ and $ W_{LTA}$, thus grant our model with more diversity. By stacking several aggregation layers, we are able to observe patterns in a large area, i.e. k-hops. The final $X^{'}$ is the node embeddings in our framework.

\subsubsection{Network Decoding}
We reconstruct the input social networks by computing the local pairwise proximity between the nodes in the network. Specifically, for each undirected edge $e_{i,j}$, we define the reconstructed links as:
\begin{equation}\label{eq:embedding}
    \hat{y}_{i,j}=\frac{1}{1+exp(-x^{'T}_i\cdot x^{'}_j)},
\end{equation}
where $x^{'}$ is a node feature vector in the network embedding space. Eq.~\ref{eq:embedding} maps the deep node embeddings $X^{'}$ to a connection matrix $\hat{Y}=\{\hat{y}_{i,j}\}$ where each element ranges from 0 to 1 and the larger value indicates the strong social connections between users.


\subsection{The Proposed Framework - {\m}}
Now, we introduce our GraphUIL model in Fig.~\ref{fig:lossrecord}. The network embedding is first applied with an encoding network which has several MSA layers equipped with a skip connection strategy~\cite{xu2018representation}. Then, a network reconstruction is conducted to regulate node embedding process. Eventually, user identity linkage across networks is built upon on the node embeddings with an MLP network. The whole learning process is controlled by three objective terms, 1) global network topology preservation; 2) local network topology preservation; and 3) cross-network mapping.

\noindent\textbf{Global Network Topology Loss}. To preserve the global network topology, we minimize the weight difference on all edges between the input and reconstructed networks.
\begin{equation}
\begin{split}
    \mathcal{L}_{global}=&\sum_{i,j}^nb_{i,j}(y_{i,j}-\hat{y}_{i,j})^2=||(Y-\hat{Y})\odot B||.
\end{split}
\end{equation}
Here, $Y$ represents the connection patterns in the social networks and we set it to be the same as the adjacent matrix $A$. $B=\{b_{i,j}:[1,0]\}$ is the sampling matrix. Due to the sparsity of social networks, preserving all unlinked edge patterns, i.e. $y_{i,j}=0$, might lead to poor performance of network embedding. Therefore, we use the negative sampling to select a number of unlinked edges and make sure the distributions of linked and unlinked edges are balanced. Our sampling strategy follows that in ~\cite{mikolov2013distributed}.

\noindent\textbf{Local Network Topology Loss}. We adopt the first-order proximity~\cite{wang2016structural} to capture the local structure. With the supervised structural information, i.e. 1-hop neighborhood connections, this term constrains the similarity of the node embedding features of a pair of users. The loss function is defined as:
\begin{equation}\label{eq:localloss}
\begin{split}
    \mathcal{L}_{local}=\sum_{i=1}^n \frac{1}{|\mathcal{N}_i|}\sum_{j\in\mathcal{N}_i}||x^{'}_i-x^{'}_j||^2_2,
\end{split}
\end{equation}
where $|\mathcal{N}_i|$ is the number neighbourhood nodes of $v_i$. Eq.~\ref{eq:localloss} generalizes the idea of Laplacian Eigenmaps~\cite{belkin2003laplacian} and drives nodes with similar embedding features together.

\noindent\textbf{Cross-network Mapping Loss}. Given a nodes pair, $(v_i,v_k)$, from two social networks separately and their embedding features, $(x^{'}_i,x^{'}_k)$, we intend to learn a mapping function $f$ if these two nodes belong to an identical user. Here, we use an MLP to capture this non-linear mapping relationship across the embedding spaces. The loss is:
\begin{equation}
    \mathcal{L}_{match}=\sum_{(v_i,v_k)\in U}||f(x^{'}_i;\Theta)-x^{'}_k||^2_2.
\end{equation}
$\Theta$ is the learnable weight parameters of MLP.

Finally, with $k=2$ social networks (SN), the total loss for  GraphUIL is a weighted combined loss,
\begin{equation}
    \mathcal{L}=\alpha\sum_i^k\mathcal{L}^{SN_i}_{global}+\beta\sum_i^k\mathcal{L}^{SN_i}_{local}+\mathcal{L}_{match},
\end{equation}
where hyperparameters $\alpha$ and $\beta$ balance the network topology embedding and user identity prediction.


\section{Experiments}
\subsection{Datasets}

We construct a real-world dataset~\footnote{We will make the dataset public for the research community.} with two kinds of social networks, Instagram and Twitter. We first obtain the seed users through an online personal information aggregation site named \texttt{about.me}~\footnote{http://about.me}. In \texttt{about.me}, people can explicitly build the profiles by adding their accounts in major social networking websites such as Facebook, Twitter, Instagram, etc. We can obtain the ground truth of linkage information for user identities directly through the information provided by users. For a specific user, we obtain the social network structure by crawling the information via the provided URLs. The crawling data were processed by removing nodes with a low degree (node degree $<5$). The statistics of social networks of the dataset are shown as in Tab.~\ref{tb:GraphProper}. Finally, we have a total of 496 users that have the ground truth of both accounts in Instagram and Twitter.


\begin{table}[t]
\small
\centering
\caption{Statistics of Instagram and Twitter datasets.}
\label{tb:GraphProper}
\begin{tabular}{|c|c|c|c|c|}
\hline
\multirow{2}{*}{Datasets} & \multicolumn{4}{c|}{Network Statistics}   \\ \cline{2-5}
                         & \#Nodes                &\#Edges         & AvgDegree                &Sparsity            \\ \hline
Instagram                     & $5864$         & $21388$    & $7.295$         & $6.2e-4$              \\ \hline
Twitter                    & $2124$         & $8413$    & $7.922$         & $1.9e-3$              \\ \hline

\end{tabular}
\vspace{-1em}
\end{table}
\subsection{Experimental Settings}
To evaluate whether or not the predicted anchor linkages truly reflect the cross-network relationships, we design a classification task based on the learned node embeddings and their mapping functions $f$. Specifically, the observed anchor linkages are divided into 3 groups: training data(60\%), validation data(10\%) and testing data(30\%). Each group contains the paired node embeddings $(x_i,x_j)$, where $x_i$ is from social networks $\mathcal{G}_1$ and $x_j$ from $\mathcal{G}_2$. We first project $x_i$ from $\mathcal{G}_1$ to $\mathcal{G}_2$ via $f(x_i)$ and then concatenate the node embeddings as $[\phi(x_i)\oplus x_j]$. These node vectors are the true user identity links in our data. Follow this strategy, we create false identity links by randomly sampling unmatched node pairs and make sure the same number of true and false links. Finally, we use a single layer network as the classifier. In this experiment, 2 evaluation metrics are reported, classification and Micro-F1 scores. For all methods, we repeat the classification 10 times and statistically compare their averaged performance with the two-sample t-test.

We stacked 3 MSA layers in our model for network encoding, where parameters $W_{GTA}$ and $W_{LTA}$ has the dimension $64\times 128$ in the 1st layer and $128\times 128$ in the rest layers. Dimension of $g$ in LTA-Agg is set to $256$ and $W^{'}$ is set to $128\times 128$. The initial node features are the Node2Vec embeddings with a $64$ dimensional vector. The MLP network for matching node embeddings has two hidden layers with the same dimension $128$. We train our model with Adam optimizer by automatic propagation.
\\\\
\noindent\textbf{Compared Methods}
The representative state-of-the-art user identity linkage algorithms
are listed as follows:
\begin{itemize}
    \item PALE~\cite{man2016predict}. A supervised framework employs embedding-based network features and constructs the cross-network extension which considers user identity links as a part of the embedding networks.
    \item FRUIP~\cite{zhou2018structure}. Unsupervised learning considers friends relationship. First, the friend feature vector is extracted with the network neighborhood patterns, and then compute their similarities for linkage prediction.
     \item Node2Vec~\cite{grover2016node2vec}. This framework learns low-dimensional network representations by observing different patterns of neighborhood connections. It simulates biased random walks on nodes.
     \item Node2Vec+MSA. Applying MSA layers with Node2Vec embeddings. A variant of our model without any network reconstruction loss, i.e. both $\alpha$ and $\beta$ equal to 0;
     \item GraphUIL w/o local. A variant of our proposed model. The local topology embedding loss is removed, i.e. $\alpha=0$.
     \item GraphUIL w/o global. A variant of our proposed model by removing global topology embedding loss, i.e. $\beta=0$.
\end{itemize}

\begin{table}[t]
\small
\centering
\caption{Performance of User Identity Linkage Prediction.}
\label{tb:predresult}
\begin{tabular}{|c|c|c|}
\hline
\multirow{2}{*}{Methods} & \multicolumn{2}{c|}{Instagram \& Twitter} \\ \cline{2-3}
                         & Accuracy                              &F1-Score            \\ \hline
PALE                     & $0.564\pm0.006$               & $0.510\pm0.012$              \\ \hline
FRUIP                    & $0.518\pm0.021$               & $0.436\pm0.030$              \\ \hline
Node2Vec                   & $0.630\pm0.009$                & $0.549\pm0.059$             \\ \hline
Node2Vec+MSA                   & $0.665\pm0.011$                & $0.603\pm0.017$             \\ \hline
GraphUIL w/o local                   & $0.710\pm0.015$               & $0.716\pm0.014$             \\ \hline
GraphUIL w/o global                   & $0.682\pm0.020$                & $0.682\pm0.020$              \\ \hline
GraphUIL                   & $\mathbf{0.754\pm0.018}^*$                & $\mathbf{0.751\pm0.017}^*$              \\ \hline

\end{tabular}
\begin{tablenotes}
\item $\mathbf{^*}$ stands for significance with $P<0.05$
\end{tablenotes}
\vspace{-1em}
\end{table}

\subsection{User Identity Linkage Performance}
We present the prediction results in Tab.~\ref{tb:predresult}. It's worth noting that we set $\alpha=10$ and $\beta=1$ in our model according to the grid searching results and put the discussion in the section of parameter analysis. For the task predicting true or false user identity linkages between Instagram and Twitter networks, our model has a significantly better performance (t-test with $P<0.05$) than other baseline methods. It achieves an averaged accuracy at $75.4\%$ with the F1-score $0.751$. Since all the baseline methods require a network embedding before network matching, their performance is largely influenced by the embedding algorithms. The best baseline method is Node2Vec ($63\%$ accuracy and $0.55$ F1-score), which is also our initial state of node features. By adding our proposed multi-stage aggregator layers, performance is improved with $66\%$ accuracy and $0.603$ F1-score. In addition, we regulate our model with reconstruction terms and observe a significant improvement than no regulated or partial regulated models. For example, compared to GraphUIL w/o local, the prediction is improved by approximate $4\%$ in accuracy when we consider local connection patterns. Besides, the global reconstruction term contributes more with increased $7\%$ accuracy when comparing GraphUIL w/o global with GraphUIL. All those results prove the effectiveness of our proposed semi-supervised graph learning model for user identity linkage prediction.

\subsection{Impact of local and global aggregators in MSA.}
\begin{table}[t]
\small
\centering
\caption{Impact of different elements in MSA.}
\label{tb:MSAelement}
\begin{tabular}{|c|c|c|}
\hline
\multirow{2}{*}{Aggregator} & \multicolumn{2}{c|}{Instagram \& Twitter} \\ \cline{2-3}
                         & Accuracy                              &F1-Score            \\ \hline
GTA-Agg Only                  & $0.672\pm0.020$               & $0.560\pm0.094$             \\ \hline
LTA-Agg Only                   & $0.710\pm0.008$                & $0.641\pm0.021$              \\ \hline
MSA                   & $\mathbf{0.754\pm0.018}^*$                & $\mathbf{0.751\pm0.017}^*$              \\ \hline

\end{tabular}
\end{table}

Since we use the MSA for network embedding, how each stage contributes to the user identity linkage prediction is unknown. We exam their impacts on our model and report results in Tab.~\ref{tb:MSAelement}. Both GTA-Agg and LTA-Agg show significant positive effects in our analysis. Without LTA-Agg, model performance drops around $0.08$ in accuracy and $0.2$ in F1-score. Meanwhile, without GTA-Agg, the model has less severity in accuracy decrease than missing LTA-Agg but still has $0.04$ drops in accuracy and $0.1$ in F1-score. From results, we postulate that the local and global awareness aggregators provide complementary information to each other and the local node information has more impacts than global topological feature when matching two social networks.

\subsection{Parameter Analysis}
\begin{figure}[t]
\center
    \includegraphics[width=1\linewidth]{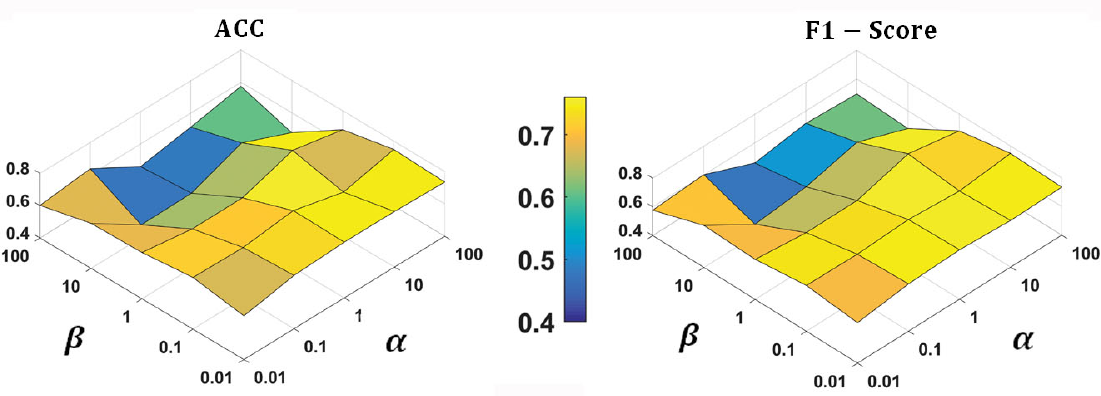}
    \caption{Parameter Analysis on weights controlling global and local losses.}
    \label{fig:grid}
\vspace{-1em}
\end{figure}

We investigate the impacts of hyperparameters $\alpha$ and $\beta$ for weighting global and local loss terms and how they control our model performance. We conduct grid sampling for each parameter and sampled them as $[0.01,0.1,1,10,100]$ respectively. The prediction accuracy and f1-score are presented in Fig.~\ref{fig:grid}. We can see that our model performs relatively well when $\alpha$ has a large value (greater than 1) while $\beta$ is small (less than 1). Meanwhile, we found the best performance at the point when $\alpha=10, \beta=1$. If we decrease $\alpha$ and increase $\beta$, the overall performance drops. This is because the small $\alpha$ leads to incomplete global network reconstruction, e.g. missing some of the subgraph details, while the large $\beta$ leads over smooth on local structures.

\section{Related Work}\label{sec:related}

\subsubsection{User Identity Linkage.}
A general framework for user identity linkage is a unified structure in most studies, which consists of two major phrases: network feature extraction and model construction. Generally, the node features are extracted from users' profile~\cite{malhotra2012studying,bartunov2012joint}, content~\cite{nie2016identifying,liu2014hydra} and network structures~\cite{zhou2016cross}.
However, in real big social networks, user's profile is generally unavailable due to privacy issues and his content features are sparse. Therefore, more attention has been drawn to feature extraction solely based on social network structures. 
Previous works investigate the network structures from two levels: neighborhood-based feature learning~\cite{zhou2016cross,vosecky2009user} which investigates the close friend connections and global network embedding learning~\cite{man2016predict,liu2016aligning} which assesses a user's role in all network connections. In our model, we believe both levels are indispensable and they should dynamically compensate each other in network embedding. The model construction can be divided into supervised, semi-supervised and unsupervised models. 
Previous supervised models differ in the design of prediction strategies~\cite{nie2016identifying,vosecky2009user} and semi-supervised models~\cite{zafarani2015user,zhang2015cosnet,tan2014mapping} generally add a reconstruction term to optimize the prediction learner and thus make it more general. In contrast, the unsupervised model learns identity linkages directly from user features by weakly aligning them through progressive procedures~\cite{liu2013s}. However, few papers worked on this model because labeled matching pairs can be easily acquired from user-self posting websites.

\subsubsection{Graph Neural Networks.}
More recently, many attempts have been made to design and apply neural networks effectively and efficiently for learning on graph structure data. Gori et al.~\cite{gori2005new} present an extended recursive neural network that copes with graphs by aggregating topological information of the neighborhood structure. The similar idea was introduced by Scarselli et al.~\cite{scarselli2009graph} to tackle different types of graph. Since the great success of convolutional neural networks on computer vision area, recently, there is an increasing interest to extend such model to the non-Euclidean domain. The key point of graph convolutional neural network is to generalize a localized convolutional operator on the graph domain. The designation of convolutional operators can be categorized as spectral~\cite{kipf2016semi,defferrard2016convolutional,levie2017cayleynets} and non-spectral approache~\cite{duvenaud2015convolutional,hamilton2017inductive}.

\section{Conclusion and Future Work}
In this paper, we proposed a semi-supervised learning model, GraphUIL, for social networks analysis. Depending on the network structures, it finds user identity linkages across multiple social networks in the node embedding space. For network embedding, we design a multi-stage aggregator to encode global and local network structures dynamically. The experimental result in a real dataset indicates the effectiveness of GraphUIL. As future works, we will devote to design a unified feature extraction model which combines various user features in addition to social network structures.

\newpage
\bibliographystyle{aaai}
\bibliography{refs}

\end{document}